2-21-2021

# Indonesian Journal of Applied Linguistics: A Bibliometric Portrait of Ten Publication Years

Abdul Syahid

# INDONESIAN JOURNAL OF APPLIED LINGUISTICS: A BIBLIOMETRIC PORTRAIT OF TEN PUBLICATION YEARS

Abdul Syahid
*Institut Agama Islam Negeri Palangka Raya, Indonesia*
abdul.syahid@iain-palangkaraya.ac.id

**Abstract**

Bibliometric portraits of a single journal appear to be rarely taken in the field of applied linguistics. Viewed from the angles of publication, citation, and indexation, one of the journals worth a bibliometric portrait is the *Indonesian Journal of Applied Linguistics*. Casting local and regional concerns on the global applied linguistics, the journal has ranked among the big five Open Access Journals in the Asiatic region since its foundation in 2011. Capturing a corpus of 426 documents by 824 authors from 144 organizations through two free bibliometric tools, i.e. *Publish or Perish* and *VOSviewer*, this study portrays the journal from 2011 to 2020 through the lens of *Microsoft Academic*, one of the largest yet free academic search engines and bibliographic databases. The portrait exhibits the journal's scientific productivity and quality, including the most prolific authors and their affiliations. It also depicts the co-authorship, keyword co-occurrence, self-citation, and bibliographic coupling. How some aspects such as the relative dominance of authors from the university publishing house has evolved before and after the Scopus indexation provide a more vivid portrait of the journal. It could provide not only retrospective but also prospective insights into the ongoing contribution of the journal to the big enterprise of applied linguistics.

## INTRODUCTION

The Indonesian Journal of Applied Linguistics (IJAL) is very worth a portrait from the angles of publication, indexation, and citation. Viewed from the publication, since 2011 IJAL has published its contents uninterruptedly. Until May 2021 IJAL could survive the financial problems even though it requires authors and readers to pay no fees at all for regular articles (IJAL, n.d.). Moreover, IJAL has relatively limited financial resources from the Indonesian government because its focus and scope do not belong to such national top research priorities as poverty alleviation (Wiryawan, 2014). IJAL is therefore one of the few journals, as Silver (2018) highlights, that could survive on such limited economic support. Similarly, IJAL could stand the test of time. The tenth volume in 2020 and 2021 shows that IJAL does not belong to over 6% of 67,082 academic journals published between 1950 and 2013 which could survive for three years only (Liu, Hu, Wang, & Shi, 2018). Besides, IJAL is characterized by its dynamic publication. For instance, IJAL published two issues a year from 2011 to 2016 and has published three ones a year since 2017 (IJAL, n.d.).

Another notable exception of IJAL could be seen from indexation. As regards the second angle, only three years after its inauguration, IJAL could justly be proud of being one of almost 41,500 scholarly journals included in *Scopus* (https://www.scopus.com/), one of the major players in the indexing enterprise (Elsevier, 2021; Scopus Title Evaluation Support, 2014). More importantly, *Scopus* has covered the journal's contents since not only the beginning indexation year, i.e. 2014 but also the beginning publication one, i.e. 2011 (Elsevier, 2020b). Some older journals in the same field needed longer time to be indexed in Scopus and their contents were covered from the beginning indexation year onward. For example, after 15 publication years *Journal of Language and Linguistic Studies* (JLLS) (https://www.jlls.org/index.php/jlls) by a Turkish public university was included in *Scopus* in 2019 and the contents were covered only from 2019 onward (Syahid & Qodir, 2021). The inclusion in one of the elite bibliographic databases suggests prestigious global recognition of scientific excellence at not only the journal but also national, institutional, departmental, and individual or author levels (Kwiek, 2020). That IJAL was successfully indexed in Scopus, however, is not the end of the matter. IJAL has to be still engaged in a quest for increased quality, otherwise it will be discontinued from Scopus. Between 2009 and 2020 Scopus (2020) excluded approximately 640 journals. Reasons for the exclusion from Scopus include the third angle-related indicators such as CiteScore (Elsevier, 2020a).

Nowadays citation, quality, impact, and significance are almost interchangeable in academia. Examining how they relate to one another in the global scholarly literature, Aksnes, Langfeldt, and Wouters (2019) conclude that the growing need for simpler and easier research assessment would amplify the use of citation-based indicators for assessment purposes. Based on the citation data in Scopus, since 2011 SCImago (n.d.) has put IJAL into the first rank of journal in Indonesia and the top ten list of Open Access Journals in the Asiatic region for two categories, namely *language and linguistics* in the area of *Arts and Humanities* along with *linguistic and language* in that of *Social Sciences*.

The applied linguistics itself sits at an intersection of not only the diversity of fields such as education and linguistics but also the range of contexts such as language teaching and multilingualism (Phakiti, De Costa, Plonsky, & Starfield, 2018). IJAL is worth examining indeed because of not only the trinity of its publication, indexation, and citation but also the local and regional concerns it could add to the global research trends in applied linguistics.

Standing as the Indonesian leading journal in the field, IJAL could be the main outlet for the vast linguistic panorama of Indonesia where about 10% of the world's 7,117 languages is used but over half of its 710 living languages were categorized as endangered ones (Eberhard, Simons, & Fennig, 2020). For instance, Usman and Yusuf (2020) reported the Acehnese's dehumanizing metaphors by presenting human as animals for negative meanings and as plants for positive ones. To deal with preserving the endangered languages, documenting them is among the first important things to do because the documentations could help restore the prestige and usage of the languages (Guérin, 2008). The further preservation effort could be facilitated by IJAL. For example, Perangin-Angin and Dewi (2020) explored the patterns and meanings of proverbs used in Pagu, a Non-Austronesian language spoken by approximately 3,300 people in the North Maluku province of Indonesia (Eberhard et al., 2020). As a journal recognized by many reputable indexing services, IJAL could help increase the visibility of the Indonesia's rich linguistic diversity.

Remarkably, only five years after its inauguration, the excellence of publication, indexation, and citation made IJAL one of the big five most preferred Scopus-indexed journals for publication by Southeast Asian scholars in the fields (Barrot, 2017). Southeast Asia is one of the most linguistically diverse regions in the world where 1,249 living languages could be found (Eberhard et al., 2020). The linguistically superdiverse region notwithstanding, the region is united by English, especially as the official working language of the Association of South East Asian Nations (ASEAN, 2008). The dynamics of English varieties by more than 130 million users (Bolton & Bacon-Shone, 2020) and English language teaching in the region's Expanding and Outer Circles have gained much attention, e.g. Lim (2020) and Zein, Sukyadi, Hamied, and Lengkanawati (2020). The dynamics could well be voiced by the South East Asian scholars through IJAL. For instance, Osatananda and Salarat (2020) from a Thai university, Thammasat University, found that Thai EFL teachers had a negative attitude to the Thai-accented English. IJAL therefore could be one of the major scientific outlets for increasing the research quantity and quality of Southeast Asian countries in language and linguistics.

In spite of the onward march of its publication, indexation, and citation, “the relative dominance of one institution among editors and authors” did not escape the attention of the Content Selection and Advisory Board when accepting IJAL for inclusion in Scopus (Scopus Title Evaluation Support, 2014). One way to track the composition of authors and their affiliations in IJAL is to take IJAL’s portrait of before and after the inclusion in Scopus. By so doing, this work could present a better full-length portrait of IJAL.

For about one decade IJAL has demonstrated the dynamic trinity of publication, indexation, and citation. Its contributions to the inter- and multidisciplinary enterprise of applied linguistics are unquestionably undeniable, at least at the national and regional levels. Portraying a leading national and regional journals such as IJAL could thus capture “the conversations being held, the theories being accepted, the research being carried out, and the tools being created” (Chen, Zou, & Xie, 2020, pp. 692–693). It could portray the cutting edge of a research area (Chen, Yu, Cheng, & Hao, 2019). Not only are both the microscopic and telescopic perspectives offering the interpretation of past research and the direction of future one, respectively (Plonsky & Oswald, 2015) of significance for the trinity of publisher, editors, and editorial board but also the binity of (future) readers and authors (Lei & Liu, 2019b). For the journal’s stakeholder, the portrait could show how well they have pursued the journal’s aims, i.e. “to promote a principled approach to research on language and language-related concerns by encouraging enquiry into relationship between theoretical and practical studies” (IJAL, n.d.). For the second party, the portrait would show whether IJAL is worth reading and contributing to.

A journal could be portrayed, at least, from a textual perspective such as a critical review of literature and a numerical perspective such as publication and citation structures. From the textual perspective of 416 empirical papers published from 2002 to 2019, a portrait of *Journal of English for Academic Purposes* taken by Riazi, Ghanbar, and Fazel (2020) reveals single authorship, genre-theory-related teaching learning issues in undergraduate setting, and adaptation of mixed research method. From the second perspective of 1,589 articles published in over four decades by *System*, a portrait taken by Lei and Liu (2019b) shows the journal’s top discussed topics, cited articles, cited references, and productive authors along with their affiliations.

Such a numerical perspective is also known as bibliometrics, a quantitative study on contents and information in research publications at a wide range of levels from a research area and topic to journal and researcher levels to serve different purposes such as evaluation, grant, and promotion (OECD & SCImago Research Group (CSIC), 2016). A bibliometric portrait could exhibit a more objective pattern of research trends by capturing the proliferation of big data on research publications, especially in the inter- and multidisciplinary fields (Chen et al., 2020) such as applied linguistics.

In the arena of applied linguistics, bibliometric portraits have been taken from a wide array of perspectives. From a global or topical perspective, many bibliometric portraits have been taken such as bi- or multi-lingualism in language learning between 2010 and 2019 (Lin & Lei, 2020). One of stunning bibliometric portrait of applied linguistics was taken by Lei & Liu (2019a). They show how dynamic the applied linguistics has been for about one decade only. Through the lens of the Social Science Citation Index, a Web of Science’s database (http://webofknowledge.com/), covering 10,028 papers published by 42 highly regarded journals between 2005 and 2016, the portrait reveals that some issues such as sociocultural ones and theories such as the complexity theory from computer science have colored applied linguistics, not only filled with linguistics and education. By capturing such data as top discussed topics, cited publications, and cited authors, they could show trends giving a substantial insight into the field.

Bibliometrics is also applied to measure a regional or national scientific performance in the field of applied linguistics, such as four Chinese related territories, namely China, Hong Kong, Taiwan, and Macau (Lei & Liao, 2017). One of the regions worth a bibliometric portrait is Southeast Asia whose share of worldwide languages is about 18% (Eberhard et al., 2020). Unfortunately, from 1996 to 2015 the region contributed only 1.8% of the global field publication and 0.8% of the global field citation as captured by Barrot (2017) through the lens of Scopus. To improve the scientific performance in the region, she suggested governmental support including research funding and more regional scientific collaborations.

However, less bibliometric studies have been published on a single journal in the realm of applied linguistics. One of the earliest bibliometric portrait of a single journal in the field was taken by Swales (1988, pp. 162–163) who reveals that in its 20-th anniversary the *TESOL Quarterly* (TQ) gained a high status as "a major journal" and "a journal of research and scholarship" but, surprisingly, not "an international journal" because of the North American authors' domination of TQ. On the other hand, in a bibliometric study into two journals, i.e. *Language Sciences* and "*Linguistics and Education*", Sethi (2016) concludes that both of them were international journals on the similar criteria used for *TQ*. So far no one has disputed the status of the three journals as international journals because of being published by major publishing houses and put in the first quartile of related categories by Scimago (n.d.), a Scopus-based journal ranking portal.

Another bibliometric portrait of a single top tier journal was taken by Lei and Liu (2019b) who portrayed the trends and contributions of *System*. Analysing 1,600 articles during 45 years, they assert that *System* could fulfil its mission to promote technology-enhanced and applied-linguistics-based solutions to foreign language education. *System*, noted Lei and Liu (2019b), could achieve the status as an international journal in 2010s or after three decades of publication because about 40% of total publication were contributed by authors from non-Western countries. From 1973 to 2009 such authors made scientific contributions of up to 28% in *System*.

Not only are top tier journals but also some reputable emerging journals in the field worth bibliometrically portraying. One of them is JLLS published in Turkey, the only EuroAsian contry which is uniquely not exactly developed nor developing one (Investopedia, 2019; Vinokurov & Libman, 2012). Displaying descriptive and network analyses of almost 500 articles published by JLSS from 2005 to 2019, Syahid and Qodir (2021) also found the dominance of a certain country or region, i.e. Turkish authors or institutions.

Most of bibliometric studies nowadays rely on academic search engines and bibliographic databases (ASEBDs). It was not until in 1964 Eugene Garfield introduced Journal Impact Factor in his Science Citation Index (Garfield, 2007) that bibliometric studies were carried out by manually compiling a list of scholars and their works, e.g. Cattell (1906). The two major databases with which to mine bibliographic data in applied linguistics are Scopus (e.g. Barrot, 2017) and one or more out of six databases in Web of Science such as Social Science Citation Index (e.g. Lei & Liu, 2019a) and Arts and Humanities Citation Index (e.g. Lin & Lei, 2020). Each of them, however, can be accessed with only an institutional subscription which costs "tens of thousands of dollars annually" (Shotton, 2018). Bibliometric studies mining bibliographic items from databases behind a paywall are relatively not transparent and reproducible for researchers whose institutions cannot afford such a subscription. And yet, emphasize Gusenbauer and Haddaway (2020, p. 184), "*transparency*" and "*reproducibility*" (emphasis in original) belong to the requirements of conducting systematic reviews including bibliometric ones. Fortunately, free to access but reliable bibliographic databases such as Microsoft Academic/ MA (https://academic.microsoft.com/) could reasonably be used (Gusenbauer & Haddaway, 2020; Harzing, 2019; Hug & Brändle, 2017; Thelwall, 2017). Through the lens of MA, for instance, Syahid and Qodir (2021) could take a 15-year bibliometric portrait of JLLS.

Today bibliometrics has been armed with not only at least 28 free and pay walled ASEBDs (Gusenbauer & Haddaway, 2020) but also bibliometric performance and mapping tools. The ASEBDs and tools are a dream come true for those who want to carry out bibliometric research more efficiently and effectively. One of the most useful bibliometric performance tools is *Publish or Perish*/ PoP (https://harzing.com/resources/publish-or-perish; Harzing, 2007). The free application software can be used to retrieve bibliographic data from seven ASEBDs, import external bibliographic data previously downloaded from four ASEBDs, and, finally, present the patterns of publication and citation in terms of 27 indicators such as h-index and g-index (Harzing, 2011). The retrieved data, either by PoP (https://harzing.com/resources/publish-or-perish; Harzing, 2007) or from a certain ASEBD, could then be visualized in terms of networks of authorship, citation, reference, and key word. The software packages used to create the network maps include VOSviewer (http://www.VOSviewer.com/van Eck & Waltman, 2020). More details on ASEBDs, bibliometric performance software, bibliometric mapping software, and bibliometric code libraries can be found in Moral-Muñoz, Herrera-Viedma, Santisteban-Espejo, and Cobo (2020).

The unquestionable exceptions of IJAL seen from the trinity of publication, citation, and indexation since its inception make IJAL worth bibliometrically visiting. It is also interesting to pay attention to the “internationality” of IJAL after the Scopus inclusion. Not only has bibliometric work on single journals in the field of applied linguistics been rarely performed but also tended to focus on established journals by big publishers, e.g. TQ (Swales, 1988). Moreover, most of the bibliometrics portraits were taken behind pay walled ASEBDs, e.g. Scopus (Lei & Liu, 2019b).

Portraying how IJAL contributed to the field during its (first) 10-year publication year along with paying particular attention to the transparency and reproducibility is of great importance. Framed by the IJAL’s aim of promoting “a principled approach to research on language and language-related concerns by encouraging enquiry into relationship between theoretical and practical studies” (IJAL, n.d.) and whether “the relative dominance of one institution among editors and authors” (Scopus Title Evaluation Support, 2014) still exists after the Scopus indexation, this study provides performance and network analyses of IJAL between 2011 and 2020.

The performance analysis attempts to investigate a general feature of how IJAL evolved its own patterns of publication and citation from the first issue of first volume in 2011 to the second issue of tenth volume in 2020. Questions related to the performance analysis in the first and second publication years along with during the 10-year time span are set out as follows:

1. How were the publication and citation patterns of IJAL?
2. Who were the most prolific authors in IJAL?
3. Which were the most productive institutions in IJAL?
4. Which were articles and the article types the top-cited ones in IJAL?
5. What were the top keywords in IJAL?

Furthermore, the second analysis investigates network features of IJAL in the first and second five publication years along with during the 10-year time span as follows:

1. Co-authorship in terms of authors and organizations;
2. Self-citation in terms of authors, documents, and organizations; and
3. Bibliographic coupling in terms of authors, documents, and organizations.

The answers provide retrospective insight and yet could provide prospective one into the ongoing contribution of IJAL to the big enterprise of applied linguistics.

## METHOD

### Bibliometric data

Downloaded in the first and second weeks of January 2021, a corpus of 426 documents including a guest editorial was included in this study. They were published in IJAL from Volume 1 Number 1 (2011) to Volume 10 Number 2 (2020). The same bibliometric items could be retrieved from a single ASEBD by two bibliometric software tools as detailed in the next two parts.

### Academic Search Engine and Bibliographic Database

The ASEBD used for data mining was MA. Using a free to access ASEBD such as MA could guarantee the transparency and reproducibility of this bibliometric attempt. In addition, MA could be directly connected with both of the bibliometric performance and mapping tools as explained in Bibliometric Tools. Using the same ASEBD for the two bibliometric tools would be guaranteed to give this study high consistency. The bibliometric data retrieved from MA were also compared with those from other ASEBDs, i.e. *Crossreff* (https://www.crossref.org/), *Google Scholar* (https://scholar.google.com/), *Pubmed* (https://pubmed.ncbi.nlm.nih.gov/), *Scopus* (https://www.scopus.com/), and *Web of Science* (http:// webofknowledge.com/). As can be seen in the downloadable supplementary materials (Syahid, 2021), MA gave the best coverage among the ASEBDs.

### Bibliometric tools

As explained in Academic Search Engine and Bibliographic Database, there were two free bibliometric tools used in the data retrieval for the performance and network analyses. The first one was the latest version of PoP (https://harzing.com/resources/publish-or-perish Harzing, 2007), *Harzing's Publish or Perish (Windows GUI Edition) 7.29.3156.7695* whereas the second one was that of VOSviewer (http://www.VOSviewer.com/van Eck & Waltman, 2020), i.e. *VOSViewer version 1.6.16*. The bibliometric performance tool was mainly used to analyze the patterns of publication and citation of IJAL during its 10-year lifetime. The bibliometric mapping tool was used for visualizing connectedness of the bibliometric data in terms of co-authorship, citation, co-occurrence, and bibliographic coupling. PoP and VOSviewer thus have different but complementary abilities.

### Procedures and data analysis

The procedures and data analysis is a variation on the ones proposed by Syahid and Qodir (2021). The two software tools could access MA after providing an Application Programming Interface key freely available in the Microsoft Research APIs Portal (https://msr-apis.portal.azure-api.net/). The search queries in PoP were filled with "Indonesian Journal of Applied Linguistics" for "*Full journal name*" and "2301-9468" for "*ISSN*". The "Years" query was filled with "0"-"0" (unspecific years), "2011"-"2020" (examined publishing years), "2011"-"2015" (the first five years), "2016"-"2020" (the second five years). To increase the search precision, as can be seen in Figure 1, all of the search queries were consecutively combined, for example "*Years*": "2011-2020", "*Full journal name*": "Indonesian Journal of Applied Linguistics", and "*ISSN*": "2301-9468".

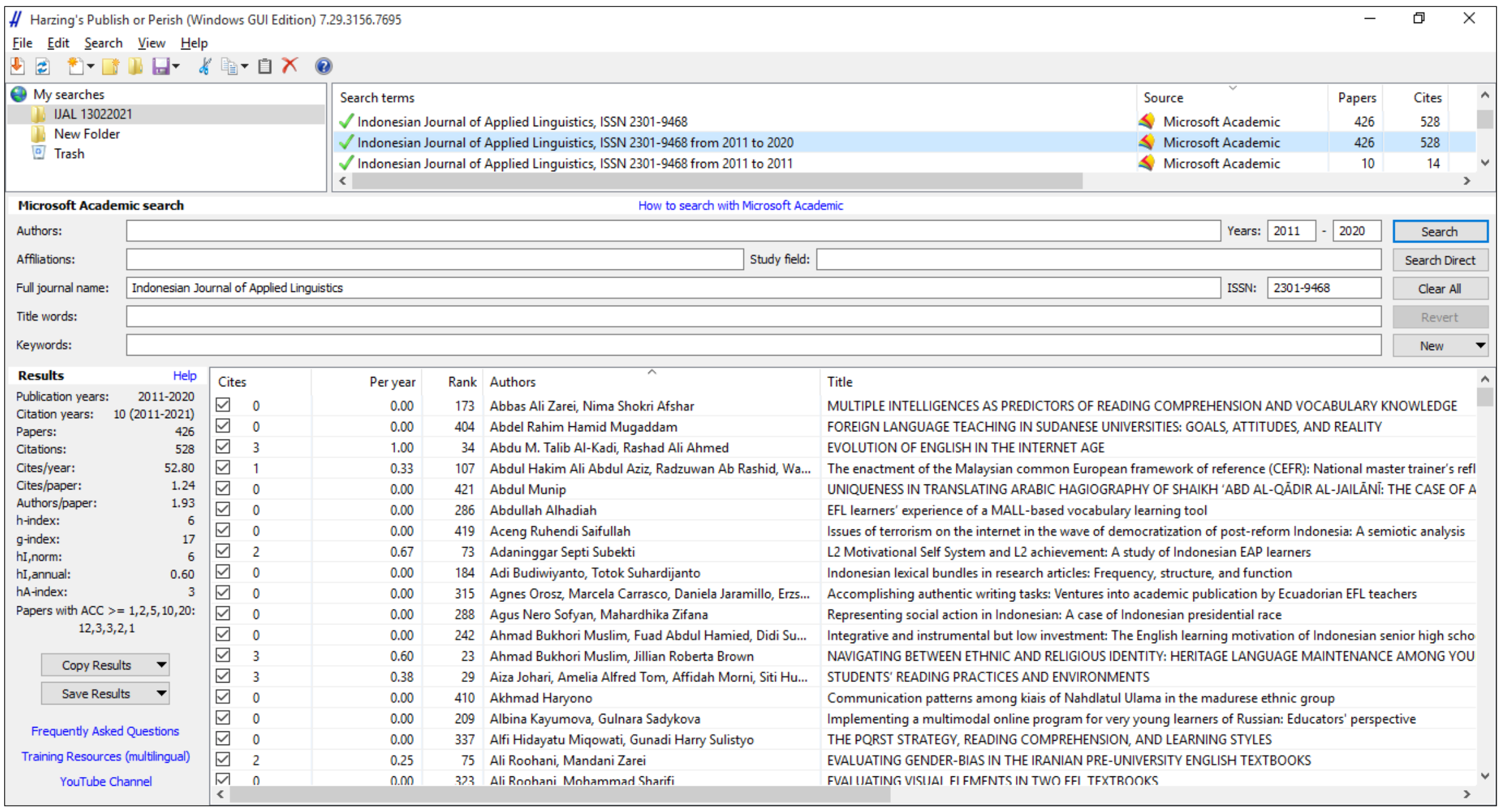

Figure 1. Search Queries in Harzing's Publish or Perish (Windows GUI Edition) 7.29.3156.7695

In the second software, a scientific landscape of IJAL was created on the basis of bibliographic data downloaded from MA. As shown in Figure 2, two out of seven search queries, i.e. "*Journal*" and "*Year*" were filled with "Indonesian Journal of Applied Linguistics" along with successively "2011-2015", "2016-2020", and "2011-2020" to increase the accuracy. The box of "*title and abstract*" was clicked. The boxes of "*Restrict to primary documents*" and "*Restrict to documents with DOI*" were checked to ensure that only documents which were published in IJAL and given a Digital Object Identifier were used to create a map. because, as shown in the online supplementary material (Syahid, 2021), nor some of the bibliographic data were retrieved from the journal's website and a few Digital Object Identifiers of the articles could be found.

After the 426 documents had been retrieved, all of the type and unit of analysis, counting method, threshold, a large number of analysis unit, and selection were set up precisely as detailed in Syahid and Qodir (2021). For a detailed review on the type of analysis and counting method in the software see Martínez-López, Merigó, Valenzuela-Fernández, and Nicolás (2018).

## FINDINGS AND DISCUSSION

The method produced a corpus of 426 documents published by IJAL during its one decade of publication. The results were reported and discussed into two clusters, namely performance and network analyses. The complete set of data are provided as downloadable supplementary materials in Syahid (2021).

### Performance Analysis

#### *Publication and citation patterns*

Figure 3 shows a total of 426 bibliometric items published during IJAL's 10 publication years, *M* = 42,6. Whereas an only decrease in publication from the previous year could be observed in 2019, two noticeable increases from the previous year could be seen, i.e. a 220% increase in 2012 and a 206% increase in 2017. Since 2017 IJAL's publication have ranged from 66 to 75 with an average paper of 72. That is why the number of publication in the second five publication year constitutes about 75% of the total publication of IJAL. From 10 documents in 2011 to 426 ones in 2020, IJAL demonstrated an Average Annual Growth Rate of 31% and a Compound Annual Growth Rate of 25%.

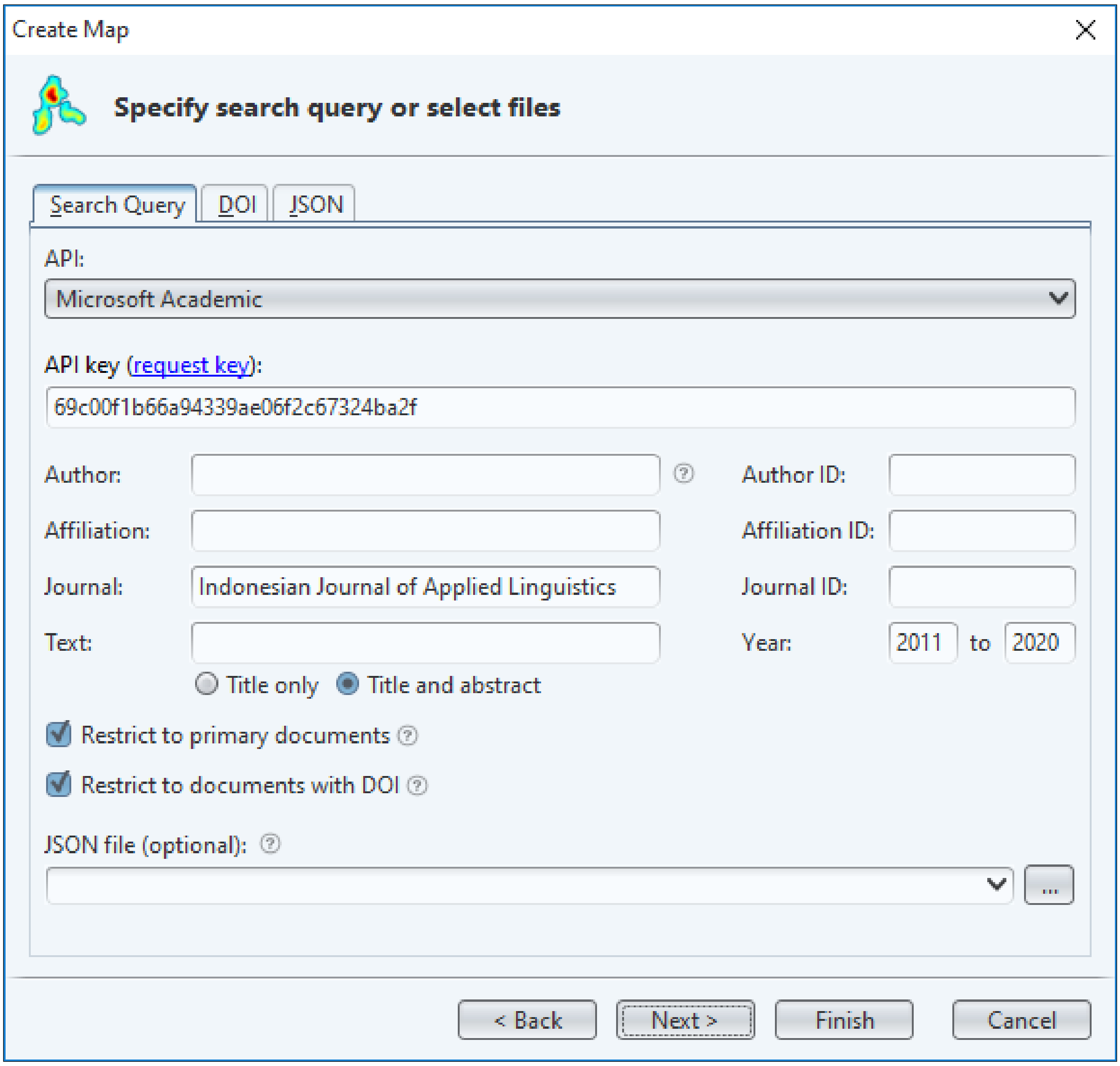

Figure 2. Retrieval Queries in VOSViewer Version 1.6.16

The noticeable boost growth in IJAL could be partially explained by the editorial change in terms of publication frequency and number of documents per issue in 2017, i.e. from two to three issues annually. Such a frequency change might result from increasing submissions after the Scopus inclusion. This could also be explained by the fact that at that time applied linguistics became wider, bigger, and faster in terms of diversity, scope, and growth (Zhang, 2020). The same boost growth could also be seen in some related journals such as *System* (Lei & Liu, 2019b) and JLLS (Syahid & Qodir, 2021).

The number of publication is a crucial factor in today's game of global academic publishing. Some indicators of significant impact in many fields such as the Scopus' CiteScore and the Web of Science's Journal Impact Factor are calculated on the basis of quantity, i.e. number of documents, and quality, i.e. number of citations (Fernandez-Llimos, 2018; Roldan-Valadez, Salazar-Ruiz, Ibarra-Contreras, & Rios, 2019). In addition, Harzing (2011) maintains that that the more journals publish documents, the bigger they have impacts on the fields. The change of publication frequency could therefore be viewed as one of concerted efforts to achieve a bigger impact.

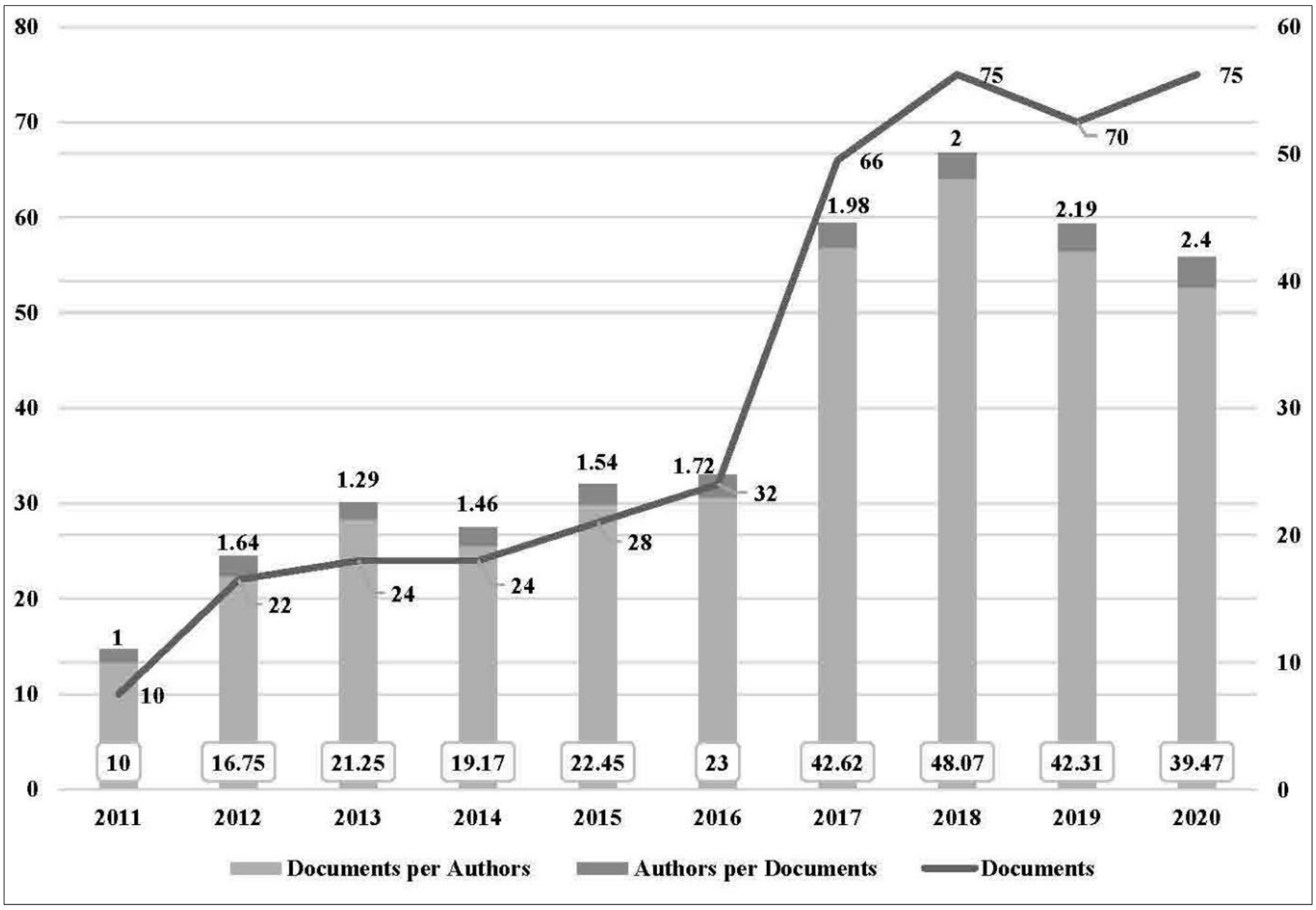

Figure 3. Publication performance of IJAL (2011-2020)

Regarding the second leg of academic publishing, Figure 4 demonstrates that during the 10 publication years 146 out of 426 documents received a total of 528 cites with an annual average of 52.8 cites, 1,2 cites per paper, and 298.55 cites per author. Over half of the total citations were received by articles published in 2012 and 2016. The citation received by 64 of 209 documents in the first five publication constitute about 52.46% of the total cites with an annual average of 50.2 cites, 0.79 cites per paper, and 124.65 cites per author. On the other hand, 82 of 317 documents between 2016 and 2020 were less cited, i.e. 251 cites with 50.2 cites per year, 0.79 cites per paper, and 124 cites per author.

The higher citation rate for earlier publication is not surprising. As is well known, the earlier articles are published, the better they have a chance of being cited (Aksnes et al., 2019; Tahamtan, Safipour Afshar, & Ahamdzadeh, 2016). To this end, some journals such as JLLS suggest that the contributors publish their submissions as preprints (Syahid & Qodir, 2021). Such an editorial policy, of which IJAL has not pursued, could help increase the visibility and citation rate of articles in IJAL.

The corpus of 426 documents in highlights a high level of citation concentration and uncitedness in IJAL. The former could be seen from six articles (less than 1.5%) with six to 109 citations accounting for slightly over half of the total citation during one decade of publication, i.e. 268 cites. The level of citation concentration is, nevertheless, not so high that the Pareto principle or 80/20 Rule simply formulated as “80% of consequences come from 20% of the causes” (Investopedia Staff, 2020) does not apply. In IJAL, 20% of the cited papers, i.e. 30 papers, accounts for 355 citations or 67%, not 80%. The principle widely applied to many fields such as computer and social sciences (Newman, 2005) appears to fail when it comes to the citation pattern in academic journals as observed in JLLS (Syahid & Qodir, 2021).

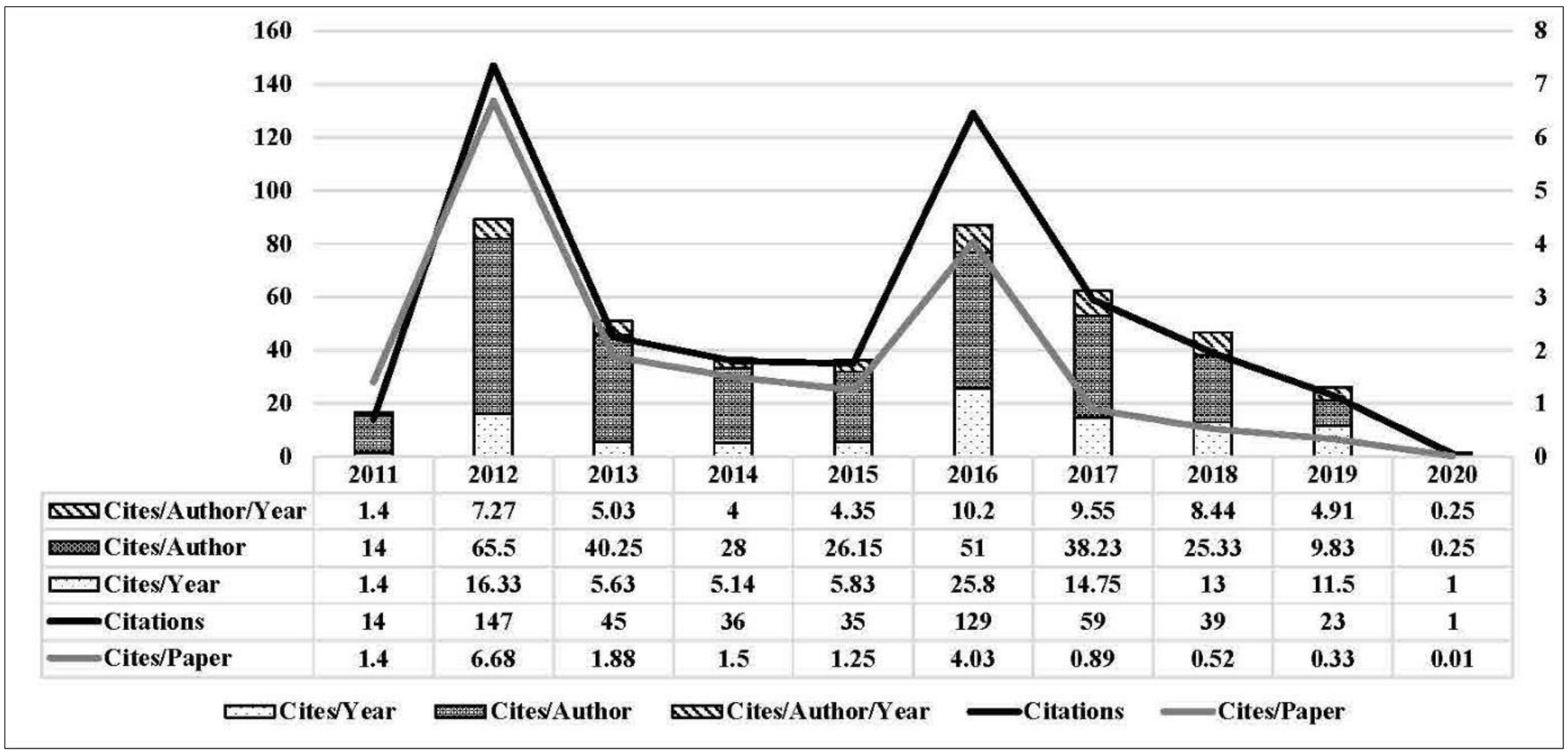

Figure 4. Citation performance of IJAL (2011-2020)

The letter could be seen from 280 documents (around 66%) in the 23 issues from Volume 1 Number 1 (2011) to Volume 10 Number 2 (2020) getting no citation. Approximately 40% of 108 documents published between 2011 and 2015 have had no citation and so have around 74% of 318 documents published from 2016 to 2020. In the field of applied linguistics, notes Harzing (2011), such high citation concentration and uncitedness are due to the waiting time for citation which is longer than that in other fields such as neuroscience and biology.

Nowadays the analysis of scientific performance is colored with more complex metrics. By adopting PoP, some of the wide range of publication and citation metrics could be presented in Table 1. Overall, the h-index of IJAL is 6, i.e. of the 426 articles, 6 articles have been cited at least 6 times or more.

Table 1. Publication and citation metrics of IJAL

| IJAL | Period | | |
|---|---|---|---|
| | 2011-2015 | 2016-2020 | 2011-2020 |
| **Individual h-index** | | | |
| h-index | 6 | 4 | 6 |
| g-index | 13 | 12 | 17 |
| hc-index | 3 | 4 | 5 |
| hI-index | 3.6 | 1.78 | 2.57 |
| hI-norm | 5 | 4 | 6 |
| **Age-Weighted (AW)** | | | |
| -Citation Rate (AWCR) | 34.33 | 66.05 | 100.38 |
| -index | 5.86 | 8.13 | 10.02 |
| -Citation Rate per Author (AWCRpA) | 22.07 | 33.37 | 55.44 |
| e-index | 10.58 | 10.72 | 14.83 |
| hm-index | 4.33 | 3.08 | 5.92 |
| hI-annual | 17.39 | 24.93 | 29.85 |
| h-coverage | 0.5 | 0.8 | 0.6 |
| g-coverage | 53.4 | 52.2 | 48.5 |
| Estimated true Citation Count (ECC) | 65.7 | 62.2 | 59.3 |
| ECC | 277 | 251 | 528 |
| **No. of Papers with Annual Citation Count per Year** | | | |
| acc1 | 3 | 9 | 12 |
| acc2 | 1 | 2 | 3 |
| acc5 | 1 | 2 | 3 |
| acc20 | 0 | 1 | 1 |
| hA | 1 | 2 | 3 |

IJAL's performance could also be seen from its g-index (Harzing, 2011) to which highly cited articles are paid more attention (Egghe, 2006). During one decade of publication IJAL had the g-index of 17, i.e. its top 17 cited articles received at least 289 cites. In line with the pattern of citation in Figure 4, not only could increases but also decreases in multidimensional metrics combining publication (quantity) and citation (quality) be identified.

### *Most productive authors*

A total of 824 authors contributed to IJAL during its ten publication years in which 155 and 669 authors were listed in the first and second halves of examined period, respectively. In Table 2 the top ten prolific authors between 2011 and 2020 along with both of five publication years were sorted on the basis of documents (D) and total link strength (TLS), i.e. "the total strength of the co-authorship links of a given researcher with other researchers" (van Eck & Waltman, 2020, p. 6). Sukyadi was not only the all-time most productive but also the only author who could stay in the top ten lists in the first and second halves of examined study period. Although Widiati was ranked among the top ten prolific authors from 2016 to 2020 only, she was the most actively involved in collaborative projects published in IJAL. Only two out of the top ten prolific authors in the 2011-2015 list had no links. To determine exactly how collaboration affects production, further work is certainly required.

Table 2. Most prolific authors in IJAL

| IJAL 2011-2020 | | | IJAL 2011 2015 | | | IJAL 2016 2020 | | |
|---|---|---|---|---|---|---|---|---|
| Author | D | TLS | Author | D | TLS | Author | D | TLS |
| D. Sukyadi* | 7 | 13 | J. S. Barrot | 3 | 1 | U. Widiati | 5 | 12 |
| N. S. Lengkanawati* | 6 | 9 | D. Sukyadi* | 2 | 4 | F. A. Hamied* | 5 | 10 |
| E. Kurniawan* | 6 | 8 | E. Emilia* | 2 | 4 | D. Sukyadi* | 5 | 9 |
| U. Widiati | 5 | 12 | N. Y. Moecharam* | 2 | 4 | E. Kurniawan* | 5 | 8 |
| E. Emilia* | 5 | 10 | S. M. Ihrom* | 2 | 4 | N. S. Lengkanawati* | 5 | 6 |
| F. A. Hamied* | 5 | 10 | A. Roohani | 2 | 2 | W. Gunawan* | 5 | 6 |
| W. Sundayana* | 5 | 9 | L. L. Amalia* | 2 | 2 | A. B. Muslim* | 4 | 9 |
| W. Gunawan* | 5 | 6 | L. Bumela | 2 | 0 | W. Sundayana* | 4 | 9 |
| S. Setyarini* | 4 | 11 | R. Adachi | 2 | 0 | A. Saukah | 4 | 8 |
| A. B. Muslim* | 4 | 9 | C. S. J. Pudin;<br>J. M. Storey;<br>L. Y. Len;<br>S. Swanto;<br>W A Din | 1 | 4 | J. Othman | 4 | 7 |

*Note:* * = affiliated with Indonesia University of Education (IUE)

The most interesting aspect of Table 2 is that most of the top ten prolific authors were affiliated with IUE, its university publishing house. A slightly similar dominance pattern was also observable in JLLS (Syahid & Qodir, 2021). It could be understood why "the relative dominance of one institution among … authors", in this case those affiliated with its university publishing house, was highlighted by Scopus Title Evaluation Support (2014). In spite of the notice, in the case of productivity, the dominance of IUE among the top ten prolific authors was higher after the Scopus inclusion than that before the Scopus inclusion. Scopus Title Evaluation Support (2014), nonetheless, views that such authorship dominance was professionally acceptable as the work by authors from IUE was apparently "generally well-written and contribute to knowledge". The productivity dominance of authors from IUE might not mean that of IUE as an institution. The next part is then to further investigate the institutional performance in IJAL.

***Most productive organization***

Although most of the top ten prolific authors were affiliated with the journal's university publishing house, such dominance could hardly be examined in terms of organizational performance. Figure 5 displays a relatively wide diversity of 144 contributing institutions in IJAL.

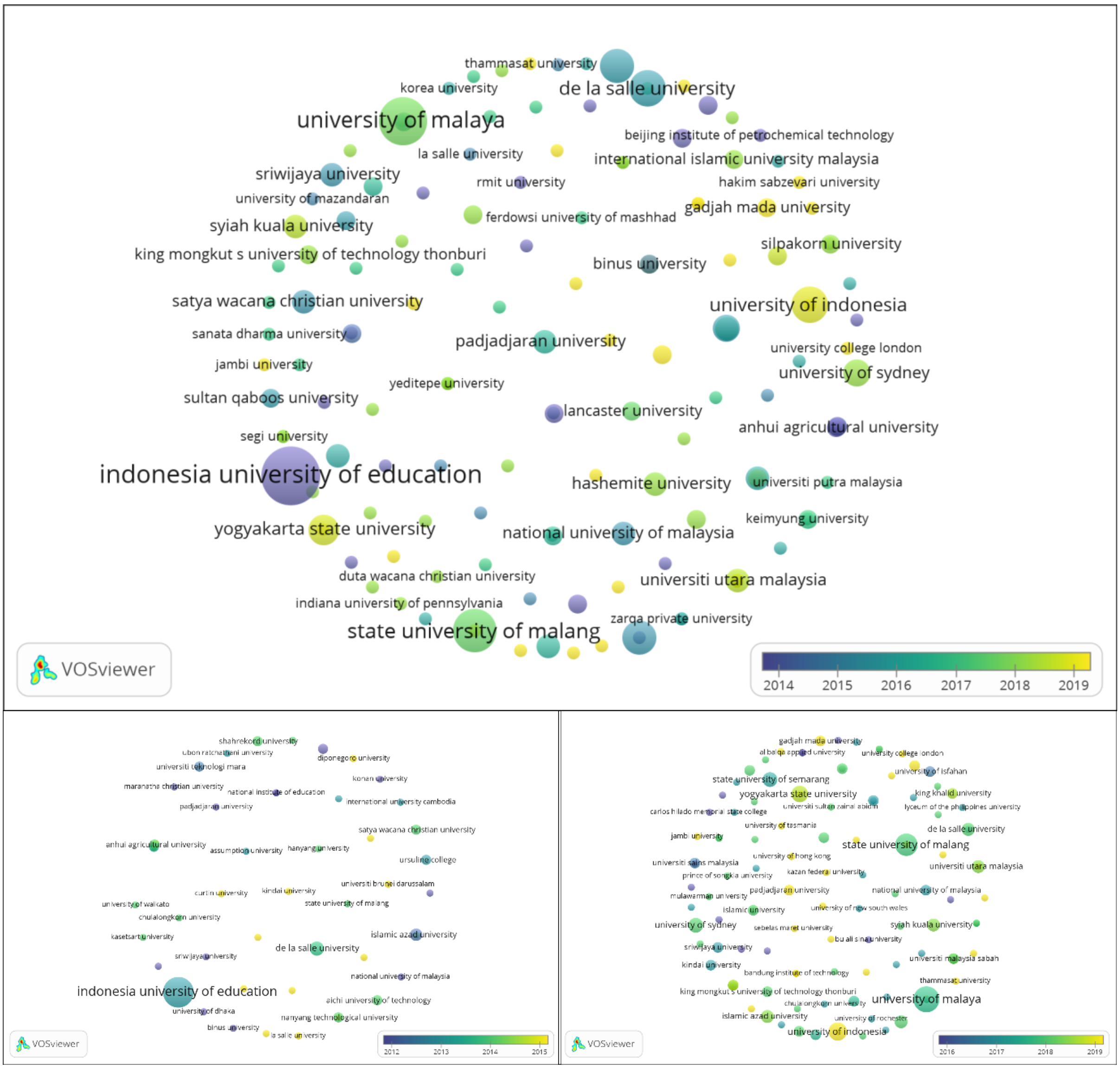

Figure 5. Organizational performance in IJAL (2011-2020)

Sorted on the basis of total documents and link strength, in the first ones they were IUE (Indonesia), De La Salle University (the Philippines), **Islamic Azad University (Iran)**, Aichi University of Technology (Japan), Anhui Agricultural University (China), Beijing Normal University (China), Nanyang Technological University (Singapore), Satya Wacana Christian University (Indonesia), Shahrekord University (Iran), and **State University of Semarang (Indonesia)**. In the second five publication years, the top ten productive institutions were University of Malaya (Malaysia), State University of Malang (Indonesia), University of Indonesia (Indonesia), Yogyakarta State University (Indonesia), **State University of Semarang (Indonesia)**, University of Sydney (Australia), **De La Salle University (the Philippines)**, Hashemite University (Jordan), **Islamic Azad University (Iran)**, and Syiah

Kuala University (Indonesia). The institutions in bold were included in the first and second five publication years but IUE was not in the second one.

The dominance of IUE among authors (and editors) in IJAL could be explained by IJAL's waiting time for getting recognized in academia. It seems possible that before having submission from authors from other institutions, IJAL had to rely on some scholars from its own university publishing house, i.e. IUE. Fortunately, before IJAL was founded, the scholars, more precisely the pioneers of IJAL, had been actively involved in the field of applied linguistics at national, regional, and international levels. Such an academic expertise enabled them to produce papers "generally well-written and contribute to knowledge" (Scopus Title Evaluation Support, 2014).

## ***Top cited articles***

The bibliometric portrait of IJAL could be taken from its top cited articles. Six out of the top ten cited articles were published between 2011 and 2015 whereas the rest was done between 2016 and 2020 (Table 3). Over 200 citations from the top two cited articles constitute about 40% of the total citations whereas all of the top ten cited articles contributed to approximately 53% of the overall citations (Figure 3). With the g- and h-indexes of 6 and 10, all of the top ten cited articles during one decade of publication belong to empirical or research-based articles focusing on educational issues.

Table 3. Top 10 cited articles in IJAL (2011-2020)

| Cites | Cites /Year | Authors | Title | Year |
|---|---|---|---|---|
| 109 | 12.11 | I Satriani, E Emilia, H Gunawan | "Contextual teaching and learning approach to teaching writing" | 2012 |
| 105 | 21 | P Purnawarman, S Susilawati, W Sundayana | "The use of Edmodo in teaching writing in a blended learning setting" | 2016 |
| 15 | 7.5 | Z Zainuddin, H. Habiburrahim, S Muluk, C M Keumala | "How do students become self-directed learners in the EFL flipped-class pedagogy? A study in higher education" | 2019 |
| 11 | 1.38 | M Islam | "English medium instruction in the private universities in Bangladesh" | 2013 |
| 8 | 1 | R Riesky | "How English student teachers deal with teaching difficulties in their teaching practicum" | 2013 |
| 8 | 0.89 | B Hermawan, N Lia | "Traces of cultures in English textbooks for primary education" | 2012 |
| 6 | 1.5 | N S Lengkanawati | "Learner autonomy in the Indonesian EFL settings" | 2017 |
| 6 | 0.86 | M W Lee | "Will communicative language teaching work? Teachers' perceptions toward the new educational reform in South Korea" | 2014 |
| 6 | 0.86 | C Viriya, S Sapsirin | "Gender differences in language learning style and language learning strategies" | 2014 |
| 5 | 1.25 | B M. Wright | "Blended learning: student perception of face-to-face and online EFL lessons" | 2017 |

These findings differ from some earlier studies (Lei & Liu, 2019a, 2019b; Syahid & Qodir, 2021). All of them noted that most of the top cited belong to conceptual articles or "just papers without data" such as theoretical and review writing (Gilson & Goldberg, 2015, p. 127). That IJAL had a younger age than the publication in those early studies did could be the reason why no conceptual article was ranked in the top ten cited papers. Another possible explanation for this is that IJAL published or received less conceptual articles despite its "encouraging enquiry into relationship between theoretical and practical studies" (IJAL, n.d.). Such top-cited-article-related topics as whether research collaboration could result in more citations could be reserved for future work.

### *Top keywords*

Based on the titles and abstracts along with unit analysis of field of study, an analysis of keyword co-occurrences was conducted by adopting VOSviewer. With a minimum number of five occurrences, Figure 6 displays 113 keywords. The top ten keywords include *psychology* (289 occurrences and link strength of 969), *linguistics* (147 and 514), *Indonesian* (107 and 419), *pedagogy* (98 and 384), *sociology* (96 and 305), *English language* (38 and 171), *curriculum* (30 and 148), *perception* (27 and 116), *foreign language* (26 and 113), and *vocabulary* (26 and 106). It also shows seven clusters of the 113 keywords.

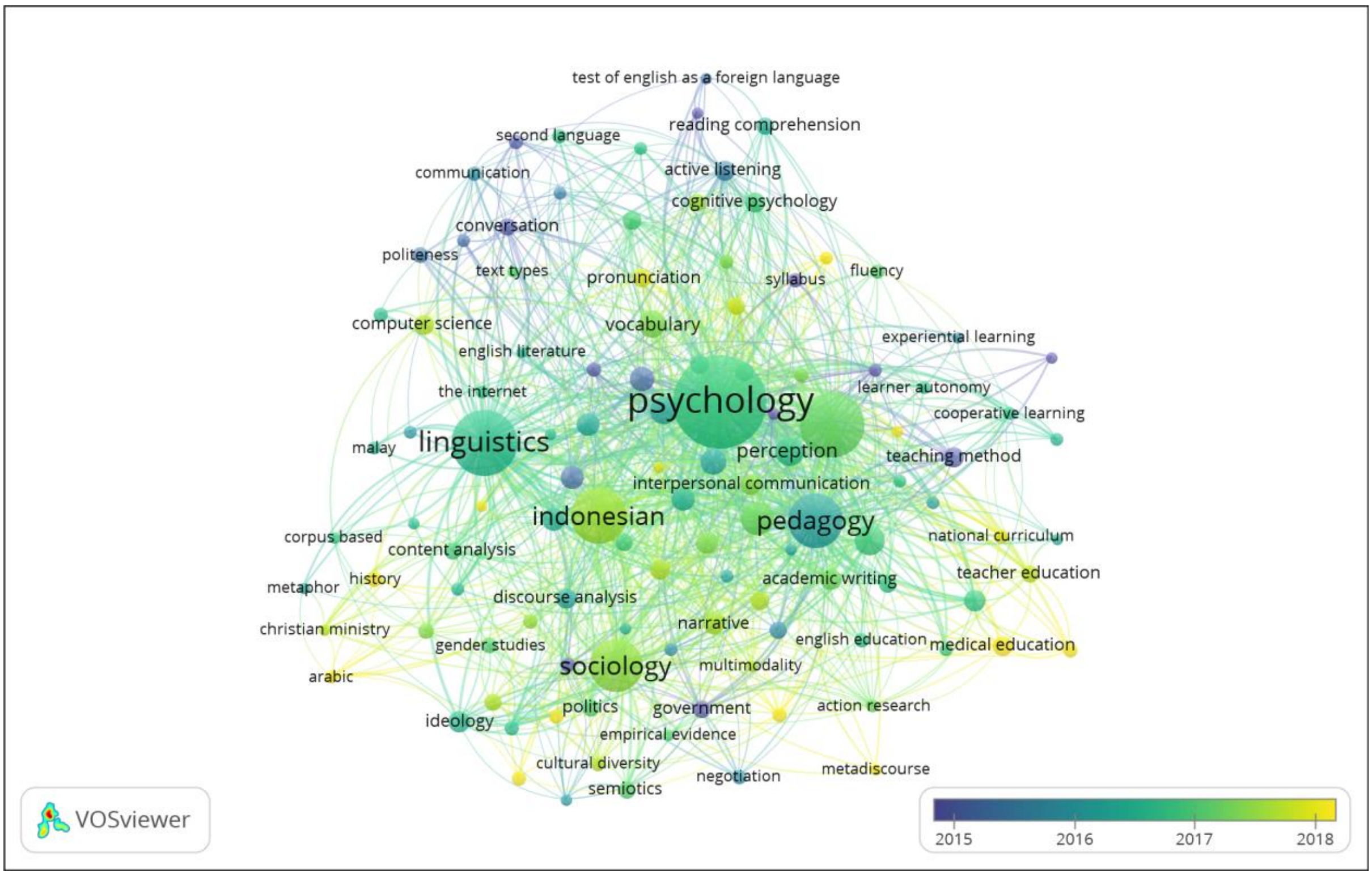

Figure 6. Keyword co-occurrences (Overlay visualization)

The results share a number of similarities with Syahid and Qodir's (2021) findings. Of 147 keywords, some similar top keywords in JLSS include *psychology*, *linguistics*, and *foreign language*. Whereas one of top keywords in JLSS was *Turkish*, one of them in IJAL was *Indonesian.* The similarities of top keywords between JLSS and IJAL could be explained by the same analysis unit of field of study in VOSviewer and the similar focus and scope of the two journals while each journal's country of publisher could be the reason why *Turkish* and *Indonesia* were included as one of the top keywords. Besides having evolved around "trinity or unity" of linguistics, psychology, and education in the realm of language teaching (Wardhaugh, 1968, p. 80), both journals have voiced their own national issues in the realm of language and linguistics. How the keywords representing "locality" would disappear as the journals become more "international" in the following decades could be deferred to future work.

## Network Analysis

### *Individual and organizational co-authorship*

As demonstrated in Figure 3, while the first five publication years have seen that sole authorship was in the majority, i.e. 1.44 authors per paper, the second ones have not, i.e. 2.1 authors per paper. The overall authorship pattern was, nevertheless, dominated by sole authorship, i.e. 1.39 papers per author (Figure 7).

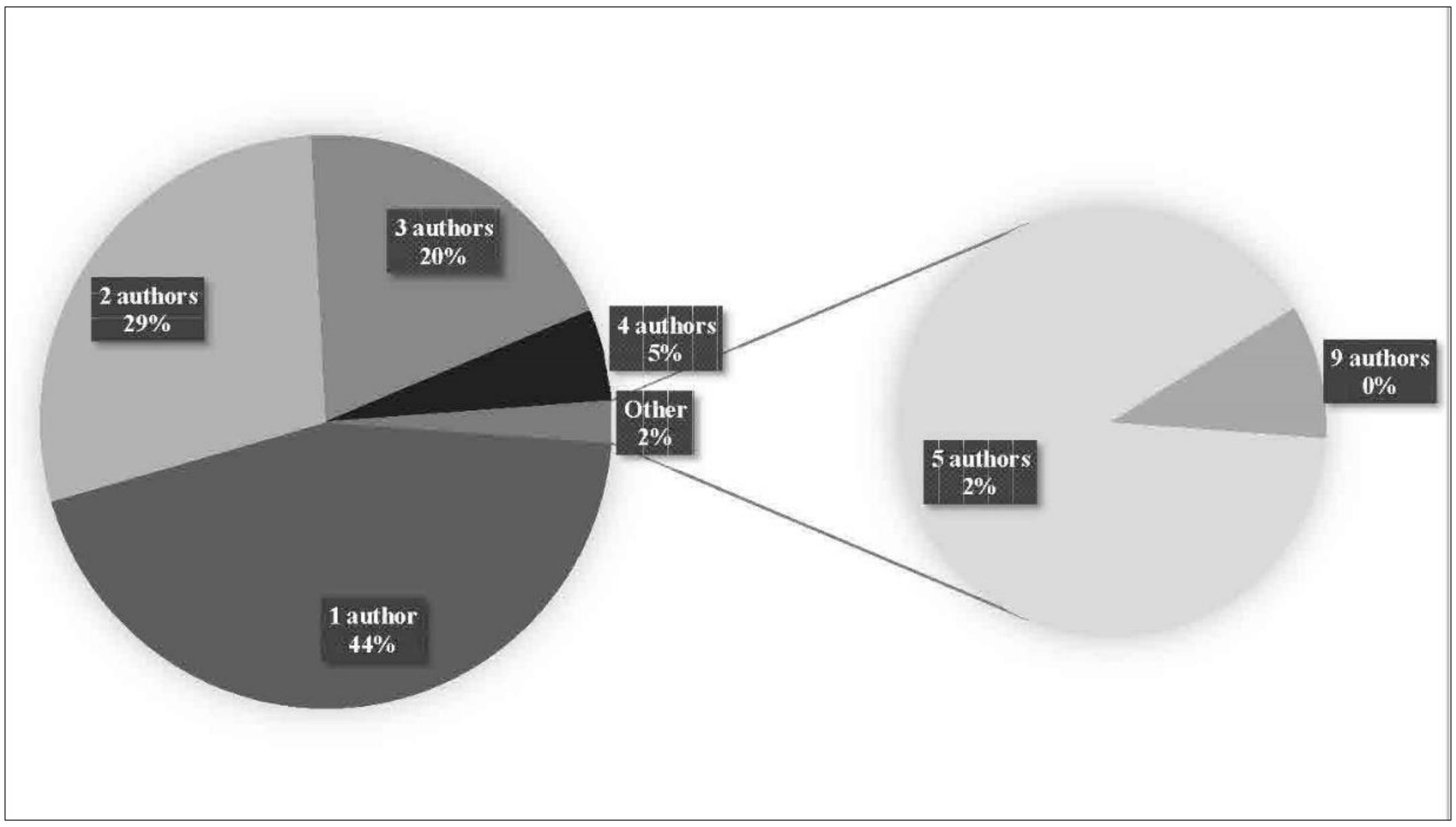

Figure 7. Authorship share in IJAL during its first decade of publication

That the sole authorship was in the majority in IJAL points to the dominance of sole authorship in language and linguistics related studies under the subject areas of Art and Humanities along with Social Sciences as observed by Harzing (2011). In the field of applied linguistics, the same pattern of authorship applied to global (e.g. Lei & Liu, 2019a) and journal (e.g. Lei & Liu, 2019b; Syahid & Qodir, 2021) levels. A relative lack of shared research agendas to applied linguistics along with the solitary and competitive nature of applied linguistics (Barrot, 2017) could explain why the sole authorship appears to be a norm even at an organizational level as shown in Figure 5.

Table 4 highlights the top ten articles by more than one authors sorted according to the number of citations. None of the articles but one was published in the first half of one decade of publication. Receiving only nine citations, the top ten papers by the number of authors had both g- and h-indexes of 2. Also found in Syahid & Qodir (2021), this apparent lack of citations received by the top ten articles with most authors could be explained by the fact that they are newly published.

Table 4. Top 10 articles by number of authors in IJAL (2011-2020)

| Cites | Authors | Author No. | Title | Year |
|---|---|---|---|---|
| 0 | D Imamyartha, S Fitriyah, Z Tasnim, A Puspa, A E Fardhani, E Wahjuningsih, S Sundari, R F Hudori, B Arya | 9 | "The efficacy of 4Cs-based reading to foster 21st-century learning competencies" | 2019 |
| 2 | C S J Pudin, J M Storey, L Y Len, S Swanto, W A Din | 5 | "Exploring L1 Interference in the writings of Kadazandusun ESL students" | 2015 |
| 0 | H Habiburrahim, Z Rahmiati, S Muluk, S Akmal, Z A Aziz | 5 | "Language, identity, and ideology: Analysing discourse in Aceh sharia law implementation" | 2020 |
| 0 | K Y Nugroho, Z Sakhiyya, M Saleh, J Mujiyanto, D Rukmini | 5 | "Exploring the constructivist mentoring program in developing EFL teacher professionalism: A qualitative approach" | 2020 |
| 0 | M W Shehzad, A Alghorbany, S A Lashari, T A Lashari, S Razzaq | 5 | "The interplay between pronunciation self-efficacy sources and self-efficacy beliefs: A structural equation modeling approach" | 2019 |
| 0 | M W Shehzad, R Ahmed, S Razzaq, A Akhtar, K Hasan | 5 | "Do reading boredom and reading boredom coping strategies predict reading comprehension performance? An empirical investigation of Saudi EFL learners" | 2020 |
| 3 | N A Drajati, L Tan, S Haryati, D Rochsantiningsih, H Zainnuri | 5 | "Investigating English language teachers in developing TPACK and multimodal literacy" | 2018 |
| 2 | S Setyarini, A B Muslim, D Rukmini, I Yuliasri, Y Mujianto | 5 | "Thinking critically while storytelling: Improving children's HOTS and English oral competence" | 2018 |
| 2 | S H Ting, E Marzuki, K M Chuah, J Misieng, C Jerome | 5 | "Employers' views on importance of English proficiency and communication skill for employability in Malaysia" | 2017 |
| 0 | S Madya, H Retnawati, A Purnawan, N H P S Putro, K. Kartianom | 5 | "The range of TOEFL scores predicted by TOEP" | 2020 |

### *Textual, organizational, and individual self-citation*

Figure 8 highlights how frequently the 426 documents (A), 144 contributing organizations (B), and 678 authors (C) cited one another in IJAL from 2011 to 2020. Most of the items of three analysis units were not connected.

At a textual level, the 426 documents made 416 clusters of which only ten were connected to each other, i.e. nine clusters of two documents and one cluster of three documents. The largest set of three connected documents consists of an article by Wright (2017) citing two articles by Ginosyan and Tuzlukova (2015) and Pollard (2015).

At an organizational level, the 144 organizations made 138 clusters of which four were connected at each other, i.e. two clusters of three organizations and two clusters of two organizations. One of the largest set of three connected organization comprises University of Malaya whose authors citing documents by authors affiliated with Anhui Agricultural University and Nanjing Normal University.

At an individual level, the 678 authors made 651 clusters. There were only eight connected clusters, i.e. three clusters of six authors, three clusters of four authors, one clusters of three authors, and one cluster of two authors. One of the largest set of connected authors comprises A Lieungnapar, D Kristina, H Hariharan, N Hashima, R W Todd, and W Trakulkasemsuk. Taken together, these findings highlight a relatively low self-citation rate in IJAL.

These findings correlate fairly well with Syahid and Qodir (2021). A reasonable explanation for the relative low citation rate in IJAL and JLSS may be that contributors to both journals consider the articles previously published are not scientifically relevant nor influential enough in their work. In the big arena of applied linguistics, most seminal work is published in top-tier or more established journals such as TQ in which self-citation could more easily be found (Swales, 1988). According to Merigó, Pedrycz, Weber, and de la Sotta (2018), self-citation is commonly found in most of journals and could help increase their bibliometric performance. That is why even some editors of highly regarded journals explicitly state that self-citation is a must for the paper acceptance (Mahian & Wongwises, 2015). If such a manually or technologically-aided unethical practice is detected, nevertheless, some elite bibliographic databases such as Scopus (Holland, Brimblecombe, Meester, & Steiginga, 2019) and Web of Science (Oransky, 2020) will exclude the journals from their indexation. The low citation rate could therefore be looked on the bright side.

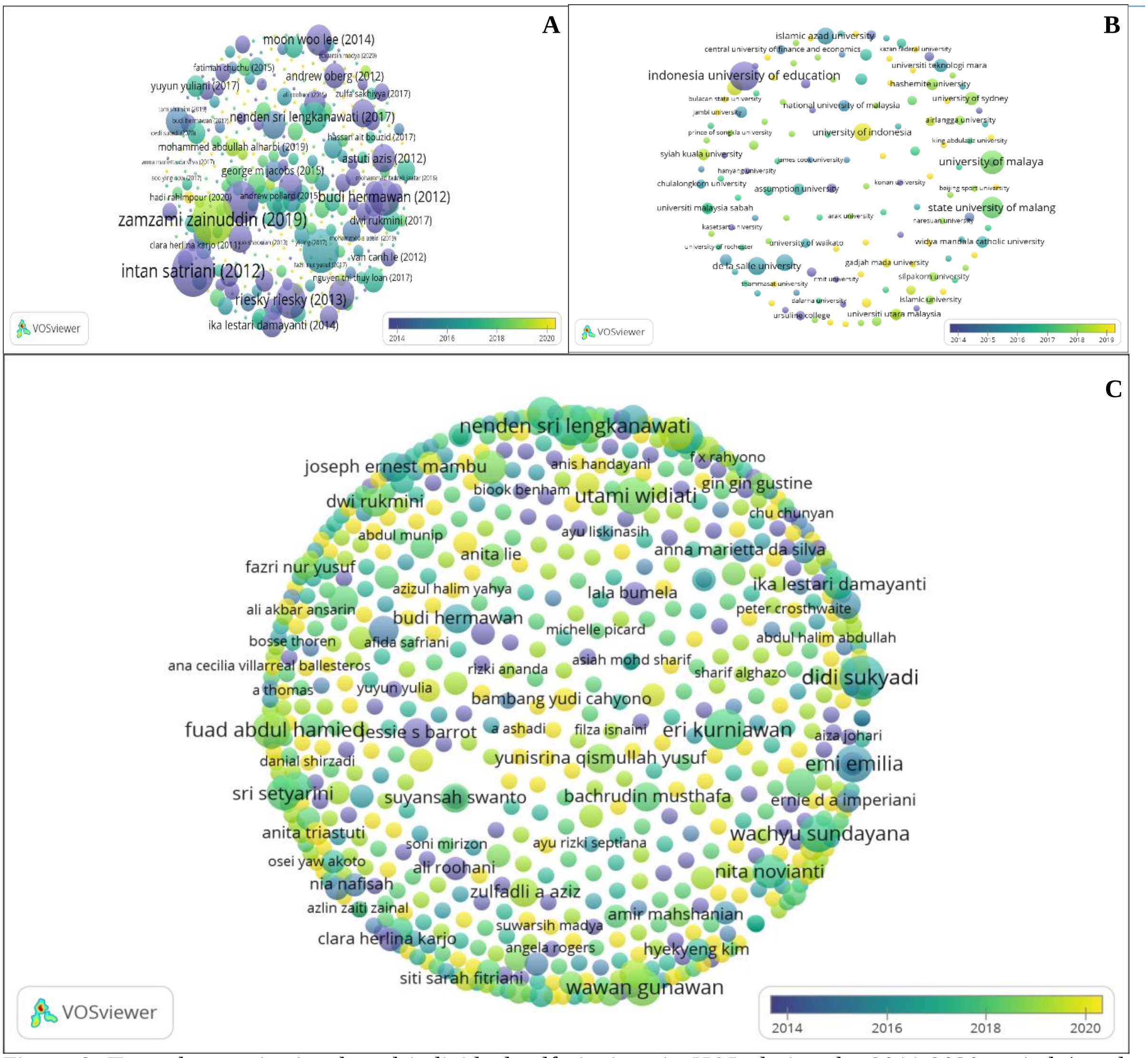

Figure 8. Textual, organizational, and individual self-citations in IJAL during the 2011-2020 period (overlay visualization)

### *Textual, individual, and organizational bibliographic coupling*

The last aspect of network analysis is concerned with the shared references. Figure 9 presents the connectedness of the documents in terms of their shared references with four units of analysis, i.e. 144 contributing organizations (A), 426 documents (B), and 678 authors (C) shared references in their work published by IJAL from 2011 to 2020. As in the self-citation, most of the items of three units of analysis were not connected to each other either.

Firstly, at a textual level, the 426 documents made 309 clusters. Only 16 clusters had a connectedness ranging from two to 24 documents having same references. Secondly, at an individual level, the 678 authors made 488 clusters. Twenty-seven clusters of authors sharing references were observable. The least set of connected authors consists of two authors whereas the largest one does 24 authors. Finally, at an organizational level, the 144 organizations made 94 clusters of organizations with which the contributing authors sharing same references were affiliated. While the largest set of connected organizations consists of 11 organizations, the least one does two organizations.

The structure of bibliographic coupling in IJAL is hardly distinguishable from Syahid and Qodir (2021) when analysing that in a 15-year-old journal in a similar field, i.e. JLLS. Such an unconnectedness might be seen as the richness of publication structure in IJAL. Because bibliographic coupling and co-citation are an intellectual portrait of a single field (Ferreira, 2018), the relative lack of connectedness of co-citation and bibliographic coupling inside IJAL reflects the wider, bigger, and faster development of applied linguistics in the past decades (Barrot, 2017), especially due to the very nature of the field as an intersection of diverse fields and contexts (Phakiti et al., 2018).

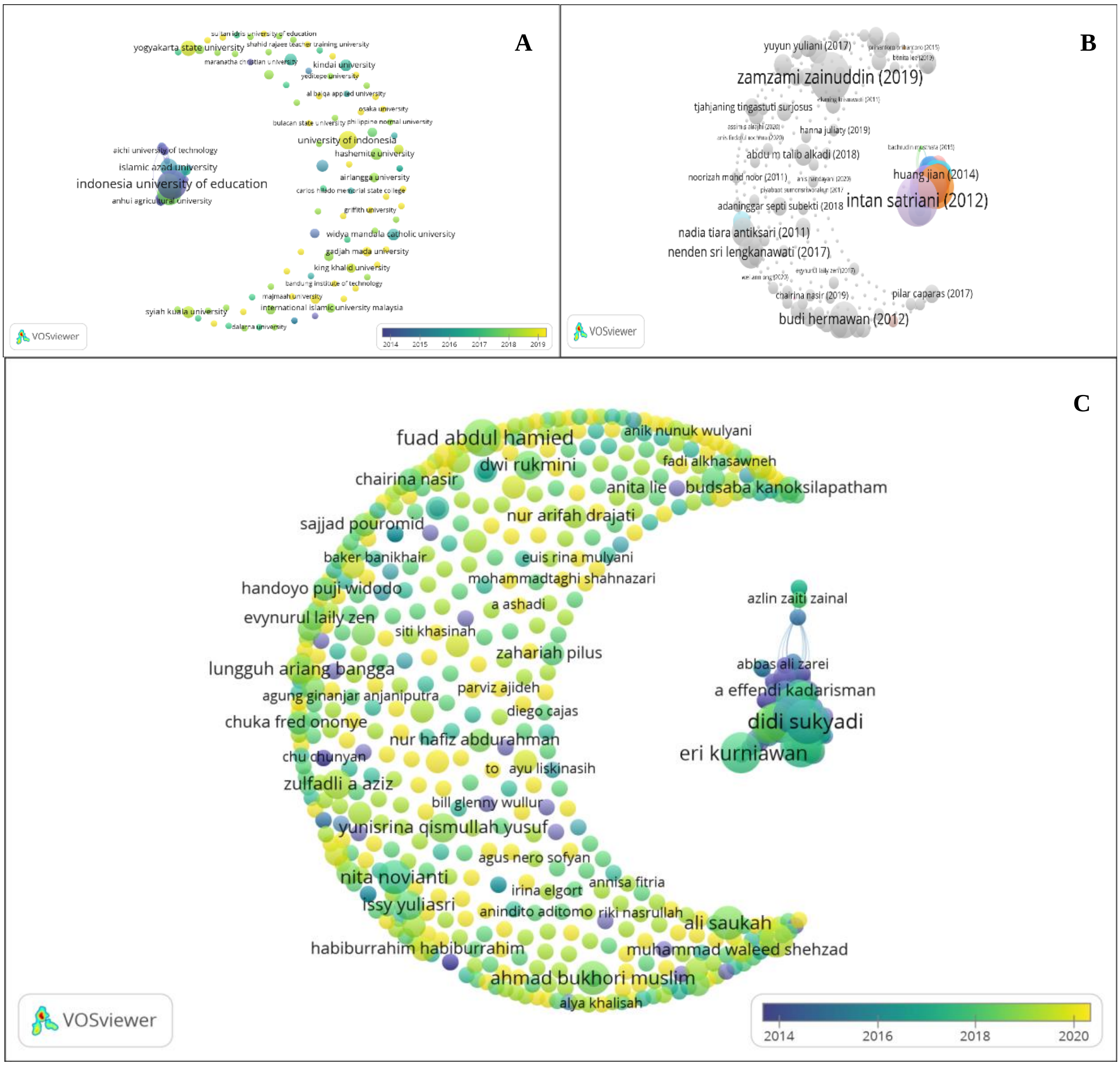

Figure 9. Textual, organizational, and individual shared references in IJAL during the 2011-2020 period (overlay visualization)

## CONCLUSION

Through the lens of Microsoft Academic, the performance and network portrait of IJAL in its first decade of publication was captured from a corpus of 426 documents by 678 authors from 144 organizations by PoP and VOSviewer. In the light of one decade of publication, the portrait exhibits how IJAL has evolved around the quaternity of psychology, linguistics, Indonesian, and pedagogy. As an emerging journal, IJAL has professionally made a significant contribution to applied linguistics by the light of national and regional concerns.

In the vast arena of applied linguistics and cut-throat competition in the global academic publishing, IJAL will always be engaged is a quest for a bigger impact and an international status. While the international status takes longer time to achieve, the greater impact could be exerted sooner by making the voice of Indonesia and Southeast Asia's rich linguistic diversity heard.

The bibliometric portrait of IJAL might be blurry in some aspects because of the limitation imposed by of the bibliographic databases and tools. Nevertheless, the portrait was taken on the tripod of best coverage, transparency, and replicability the database and tools could offer including the multidimensional perspectives such as bibliographic coupling. Also, the bibliometric portrait of IJAL was taken from the past publication and citation. Despite such a microscopic and quantitative approach, the portrait could also be used for telescoping the future development of applied linguistics to which IJAL could contribute more significantly. Another bibliometric portrait of IJAL in its two decades of publication or after IJAL applies Article Processing Charge will be a challenge.

## ETHICS COMMITTEE APPROVAL

The author confirms that the study does not need ethics committee approval according to the research integrity rules in their country (Date of Confirmation: 18/02/2021).

## ACKNOWLEDGEMENTS

Support was given by *Fakultas Tarbiyah dan Ilmu Keguruan*, *Institut Agama Islam Negeri Palangka Raya, Indonesia*.